\documentstyle[aps,psfig,float]{revtex}
\begin{document}
\draft

\title{Lattice pinning of magnetic domains in the helimagnet 
Ba$_{2}$CuGe$_{2}$O$_{7}$} 

\author{J. Chovan and N. Papanicolaou}
\address{Department of Physics, University of Crete, and 
Research Center of Crete}

\maketitle
\date{\today}

\begin{abstract}
The layered magnetic compound Ba$_{2}$CuGe$_{2}$O$_{7}$ exhibits spiral 
antiferromagnetic order thanks to a Dzyaloshinskii-Moriya (DM)
anisotropy that is allowed by crystal symmetry. Here we theoretically examine
some finer issues such as the experimentally observed lattice pinning
of the propagation vector of helical magnetic domains along the crystallographic
(1,1,0) or (1,\={1},0) direction.
We find that DM anisotropy alone would actually lead to pinning along the
(1,0,0) or (0,1,0) direction, but agreement with experiment is restored upon
including an additional exchange anisotropy that is also consistent with symmetry.
The present results also shed light on the so-called bisection rule which 
has been abstracted from experiment in presence of an in-plane magnetic field.
\end{abstract}


Ba$_{2}$CuGe$_{2}$O$_{7}$ is an essentially two-dimensional spin system that 
exhibits spiral antiferromagnetic order due to a Dzyaloshinskii-Moriya (DM) 
anisotropy \cite{1,2}.
A series of experiments, including standard magnetometry, elastic, and inelastic
neutron scattering \cite{3,4,5,6,7}, have established that the ground state
is an incommensurate spiral with period (pitch) $L\approx$ 37 lattice units
(219 \AA) at zero external magnetic field.
When a field of strength $H$ is applied along the $c$-axis the period $L=L(H)$ 
grows to infinity at $H_c\approx 2$ T and the ground state degenerates
into a commensurate spin-flop state for $H>H_c$.
Thus one obtained the first experimental realization of 
an incommensurate-to-commensurate (IC) phase transition whose gross
features were theoretically predicted some time ago by Dzyaloshinskii \cite{8}.

Ba$_{2}$CuGe$_{2}$O$_{7}$ is an insulator and its magnetic properties
may be understood in terms of localized spins. Crystal symmetry provides the
necessary input for the construction of a discrete Heisenberg Hamiltonian,
but theoretical analysis is greatly facilitated by a continuum approximation 
which leads to a certain effective field theory that is a variant of the 
relativistic nonlinear $\sigma$ model. The continuum approximation is
justified by the fact that anisotropy is sufficiently weak and thus leads
to a ground state which is a spiral with a reasonably large period.
Hence the nonlinear $\sigma$ model may safely be employed for the study
of ground-state properties and the corresponding low-energy dynamics.
The resulting theoretical picture was found to be in general agreement
with experiment \cite{3,4,5,6,7} and further clarified the precise nature
of the Dzyaloshinskii-type IC transition \cite{9,10}.
Yet, discreteness effects are important to understanding
some finer issues of experimental interest and are the main focus
of the present paper.

The unit cell of Ba$_{2}$CuGe$_{2}$O$_{7}$ is partially illustrated in Fig. 1
where we display only the magnetic Cu$^{2+}$ ions with spin $s=1/2$. 
The lattice constants are
$a=b=8.466$ \AA $ $ and  $c=5.445$ \AA. Since the Cu atoms form a perfect
square lattice within each layer, it is also useful to consider the
orthogonal axes $x, y$ and $z$ obtained from the conventional crystal axes
$a, b$ and $c$ by a $45^{\circ}$ azimuthal rotation. 
The major spin interaction between nearest in-plane neighbors
along the $x$ or $y$ axis is antiferromagnetic, while the interaction 
between nearest out-of-plane neighbors along the $z$ axis is ferromagnetic
and rather weak \cite{3}. Therefore, we are effectively dealing with a 2D
antiferromagnet defined on a square lattice with natural axes $x$ and $y$
and lattice constant $d=a/{\sqrt{2}}=5.986$ \AA.

Hence, to a first approximation, the spin dynamics is described
in terms of a 2D isotropic Heisenberg model. However, because of the
low tetragonal symmetry of this crystal (space group $P\bar{4}2_{1}m$),
the effective Heisenberg Hamiltonian may contain an interesting
combination of antisymmetric (DM) and symmetric exchange anisotropies.
In the following, anisotropy is introduced in steps of increasing
complexity, invoking the results of a complete symmetry analysis
\cite{9,10} as they become necessary.  

Experiments were initially \cite{3,4,5,6} analyzed in terms of a relatively
simple extension of the isotropic Heisenberg model defined by the Hamiltonian
\begin{equation}
\label{eq:1}
   W = \sum_{<\bf{rr}'>}\Big[ \,J\left({\bf S}_{\bf{r}}\cdot {\bf S}_{\bf{r}'}\right)
      + \bf{D}_{\bf{rr}'}\cdot \left({\bf S}_{\bf{r}} \times {\bf S}_{\bf{r}'}\right)
      \Big],	
\end{equation}
where the sum extends over all bonds $<{\bf rr}'>$ connecting any two neighboring
sites ${\bf r}$ and ${\bf r}'$ on a square lattice. Symmetry requires that the exchange
constant $J$ is the same for all bonds, while the constant vectors ${\bf D}_{{\bf rr}'}$
which account for pure DM anisotropy are of the form
\begin{eqnarray}
\label{eq:2}
{\bf D}_{{\bf rr}'} &=& (\,0,D_{\perp},\pm D_{z})\hspace{0.2cm}\textrm{for bonds along} \hspace{0.2cm}x,
\nonumber
\\
{\bf D}_{{\bf rr}'} &=& (D_{\perp},0,\pm D_{z}) \hspace{0.2cm}\textrm{for bonds along} \hspace{0.2cm}y,
\end{eqnarray}
where Cartesian components are taken along the axes $x$, $y$ and $z$ of Fig. 1, and
a sign alternation at consecutive bonds is present in the third component $\pm D_{z}$.
In fact, all available experimental evidence suggests that $D_{z}{\ll}$$D_{\perp}$
in Ba$_{2}$CuGe$_{2}$O$_{7}$.\begin{figure}
\centerline{\hbox{\psfig{figure=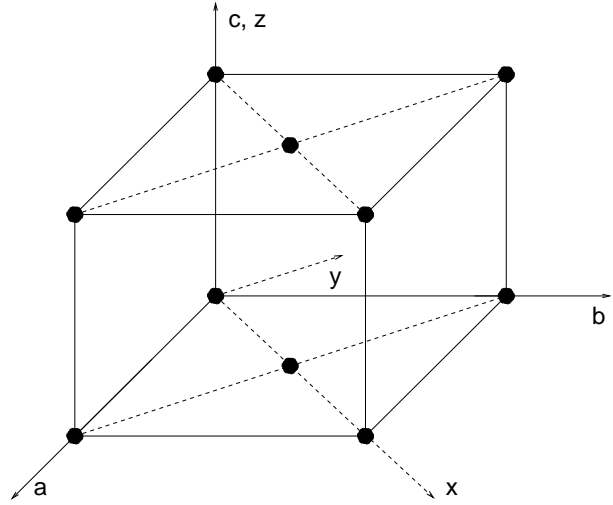,width=8.cm}}}
\vspace*{0.5cm}
\caption{Partial illustration of the unit cell of 
Ba$_{2}$CuGe$_{2}$O$_{7}$ displaying only the magnetic Cu$^{2+}$ ions, 
denoted by solid circles, which form a perfect square lattice within each layer.}
\end{figure} 
Therefore, while possible implications of a finite $D_{z}$ are studied in Refs. \cite{9,10},
we set $D_{z}=0$ throughout the present paper. Hamiltonian (\ref{eq:1}) is then written
in a completely explicit form:
\begin{eqnarray}
\label{eq:3}
 W &=& \sum_{mn} \Big[\, J\,{\bf S}_{m,n} \cdot (\, {\bf S}_{m+1,n}+{\bf S}_{m,n+1}) 
 +\, D_{\perp}(\, S^{z}_{m,n}S^{x}_{m+1,n} - S^{x}_{m,n}S^{z}_{m+1,n}\, ) 
\nonumber
\\
& &  +\, D_{\perp} (\, S^{y}_{m,n}S^{z}_{m,n+1} - S^{z}_{m,n}S^{y}_{m,n+1})\, \Big],
\end{eqnarray}
where $m$ and $n$ are integers that advance along the $x$ and $y$ axes of the square lattice
and ${\bf S}_{m,n}=(S^{x}_{m,n},S^{y}_{m,n},S^{z}_{m,n})$ is the spin operator at site ($m,n$).
At this point we invoke the classical approximation whereby spin operators are treated
as classical vectors with magnitude  ${\bf S}^{2}_{m,n}=s^{2}$. The ground state
is then described by the classical spin configuration that minimizes Hamiltonian (\ref{eq:3}).
To search for the minimum we employ the spherical parametrization
\begin{eqnarray}
\label{eq:4}
 S^{\,x}_{m,n} &=& s\,(-1)^{m+n}\sin{\Theta_{m,n}}\,\cos{\Phi_{m,n}}\,, 
\nonumber
\\
 S^{\,y}_{m,n} &=& s\,(-1)^{m+n}\sin{\Theta_{m,n}}\,\sin{\Phi_{m,n}}\,, 
\nonumber
\\
 S^{\,z}_{m,n} &=& s\,(-1)^{m+n}\cos{\Theta_{m,n}}\,,
\end{eqnarray}
where the staggering factor $(-1)^{m+n}$ reflects the antiferromagnetic nature of the basic
exchange interaction. Inserting Eq. (\ref{eq:4}) in Eq. (\ref{eq:3}) yields an energy functional
whose variation leads to a coupled system of nonlinear equations for the angular
variables ${\Theta}_{m,n}$ and ${\Phi}_{m,n}$.
We shall not write out these equations explicitly, but merely state a class 
of analytical spiral-like solutions:
\begin{equation}
\label{eq:5}
   {\Theta}_{m,n} = m\,{\gamma}_{1} + n\,{\gamma}_{2} ,\hspace{1.2cm} {\Phi}_{m,n} = -\psi ,
\end{equation}
where the \emph{constants} ${\gamma}_{1}$, ${\gamma}_{2}$, and $\psi$ are constrained only
by the curious relation
\begin{equation}
\label{eq:6}
   \tan{\psi} = \frac{\sin{{\gamma}_{2}}}{\sin{{\gamma}_{1}}}.
\end{equation}
The corresponding energy per unit cell is given by
\begin{equation}
\label{eq:7}
 w = s^{2}J \, \Big[ \, 2 - \cos{{\gamma}_{1}} - \cos{{\gamma}_{2}}
 - {\varepsilon}\left(\sin{\gamma}_{1}\cos{\psi} + \sin{\gamma}_{2}\sin{\psi}\right) \, \Big]  
\end{equation}
where the trivial constant $2s^{2}J$ amounts to removing the energy of the pure 
N\'eel 
state, and
\begin{equation}
\label{eq:8}
  {\varepsilon} = D_{\perp}/J 
\end{equation}
is a dimensionless anisotropy constant whose magnitude is left arbitrary for the moment.

The helical spin configurations constructed above are stationary points of the energy
functional for any choice of the constants ${\gamma}_{1}$, ${\gamma}_{2}$, and $\psi$ that
satisfy the constraint (\ref{eq:6}). However, the absolute minimum of the energy is achieved 
for the specific choice ${\gamma}_{1}={\gamma}_{2}=\arctan{({\varepsilon}/\sqrt{2})}$
and $\psi={\pi}/4$, or
\begin{eqnarray}
\label{eq:9}
 & \Theta_{m,n} = (m+n)\arctan{( \varepsilon /\sqrt{2})}, 
\hspace{1.2cm} 
\Phi_{m,n} = -{\pi}/4 \,,&
\nonumber
\\
\nonumber
\\
 &\displaystyle{ w = 2s^{2}J\left[\, 1 - \sqrt{1+{\varepsilon}^{2}/2}\, \right] = 
{\varepsilon}^{2} s^{2}J\left( - \frac{1}{2} + \frac{{\varepsilon}^{2}}{16} + ...\right)},&
\end{eqnarray}
which is a screw-type spiral whose propagation vector points along the crystalographic
$b=(0,1,0)$ axis, while the spin rotates in a plane perpendicular to the same axis.
An equivalent solution that propagates along the $a=(1,0,0)$ axis is also possible but will
not be given further consideration.

The preceding result is already contradicted by the experimental fact that the spiral propagates
along the $x=(1,1,0)$ or the $y=(1,\bar{1},0)$ axis \cite{3,4,5,6,7}.
To make this fact completely explicit we consider a special member of the family of solutions 
(\ref{eq:5}) obtained by setting $\psi = 0$, and thus ${\gamma}_{2}=0$ by virtue of the
constraint (\ref{eq:6}), while the remaining constant ${\gamma}_{1}$ is chosen to minimize
the reduced energy $w=s^{2}J (1 - \cos{\gamma}_{1} - {\varepsilon}\sin{\gamma}_{1})$ to
yield ${\gamma}_{1}=\arctan{(\varepsilon)}$ and
\begin{eqnarray}
\label{eq:10}
 & \Theta_{m,n} = m\,\arctan{(\varepsilon)},
\hspace{1.2cm}\Phi_{m,n} = 0 \,,& 
\nonumber
\\
\nonumber
\\
 & \displaystyle{ w = s^{2}J\left[\, 1 - \sqrt{1+{\varepsilon}^{2}}\,\right] = 
{\varepsilon}^{2} s^{2}J\left(-\frac{1}{2} + \frac{{\varepsilon}^{2}}{8} + ...\right)},&
\end{eqnarray}
which is a spiral that propagates along the $x$ axis while the spin rotates
within the $xz$ plane. Both of these facts are consistent with experiment, but the energy
of the spiral of Eq. (\ref{eq:10}) is \emph{larger} than the energy of the screw-type spiral
of  of Eq. (\ref{eq:9}) to within terms of order ${\varepsilon}^{4}$. In other words,
the pure DM Hamiltonian (\ref{eq:3}) cannot explain the observed pinning of the
propagation vector of the spiral along the $x$ or $y$ axis.

Before discussing a resolution of this apparent discrepancy, it is instructive to view
the analytical solutions given by Eqs. (\ref{eq:5})-(\ref{eq:7}) from a slightly
different perspective.Thus we attempt to first minimize the energy (\ref{eq:7}) at fixed angle $\psi$, with
${\gamma}_{1}$ and ${\gamma}_{2}$ related by the constraint (\ref{eq:6}).
The result of this minimization is written here in a perturbative form:
\begin{eqnarray}
\label{eq:11}
\sin{{\gamma}_{1}} &=&
{\varepsilon}\cos{\psi}
\left[ \, 1 - \frac{1}{2}\left(\cos{^{4}\psi} + \sin{^{4}\psi}\right){\varepsilon}^{2} + ...\right],
\nonumber
\\
\sin{{\gamma}_{2}}  &=& 
{\varepsilon}\sin{\psi}
\left[ \, 1 - \frac{1}{2}\left(\cos{^{4}\psi} + \sin{^{4}\psi}\right){\varepsilon}^{2} + ...\right],
\end{eqnarray}

and the corresponding energy per unit cell is given by
\begin{equation}
\label{eq:12}  
 w = {\varepsilon}^{2}s^{2}J \left\{ -\frac{1}{2} 
+ \frac{1}{32}\left[\,3 + \cos{(4\psi)}\,\right]{\varepsilon}^{2} + ...\right\}.
\end{equation}
This result is consistent with Eqs. (\ref{eq:9}) and (\ref{eq:10}), for $\psi=\pi/4$ and
$\psi=0$, respectively, and reinforces our earlier conclusion that the absolute minimum
of the energy is achieved for $\psi=\pi/4$ which leads to the screw-type spiral of Eq. (\ref{eq:9}).

For sufficiently weak anisotropy $({\varepsilon}{\ll}1)$ Eq. (\ref{eq:11}) may be further 
approximated by ${\gamma}_{1}\approx{\varepsilon}\cos{\psi}$ and  
${\gamma}_{2}\approx{\varepsilon}\sin{\psi}$ which are inserted in Eq.(\ref{eq:5}) to yield
\begin{equation}
\label{eq:13}
  {\Theta}_{m,n} \approx {\varepsilon}\left( m\,\cos{\psi} 
 + n\,\sin{\psi} \right), \hspace{1.cm} {\Phi}_{m,n} = -\psi .
\end{equation}
This is a class of degenerate helical configurations with common energy 
$w \approx -{\varepsilon}^{2}s^{2}J/2$ which is independent of angle $\psi$.
Eq. (\ref{eq:13}) describes a helix that propagates along the $x'$ axis of Fig. 2, which forms an
angle $\psi$ with respect to the $x$ axis, while the spin rotates in the $x''z$ plane whose intersection
with the basal plane forms an angle $-\psi$ with respect to the $x$ axis. The angle formed
by the normal to the spin plane (axis $y''$) and the spiral propagation vector (axis $x'$) is
bisected by the conventional crystal axis $b=(0,1,0)$. Therefore, to leading 
${\varepsilon} \ll 1$ approximation, theory predicts that spirals may propagate in an arbitrary direction
(angle $\psi$) as long as they conform with the ``bisection rule'' just described.

However, when the terms of order ${\varepsilon}^{4}$ are included in the energy, 
as in Eq. (\ref{eq:12}), a screw-type spiral propagating along the $a$ or $b$ axis ($\psi=\pm \pi/4$) is 
predicted to be energetically favorable. As mentioned already, this prediction is contradicted by 
experiment where spirals always propagate along the $x$ or $y$ axis ($\psi=0$ or $\pi/2$) 
in the absence of an in-plane external magnetic field. Therefore, the pure DM anisotropy included
in Hamiltonian (\ref{eq:1}) needs to be ammended.

A more general type of anisotropy is suggested by the early work of Kaplan \cite{11}, and of Shekhtman, 
Entin-Wohlman, and Aharony \cite{12,13,14}, which is usually referred to as the KSEA anisotropy. The 
latter was already invoked in Refs. \cite{7,9,10} to explain the observed distortion of the spiral from
its ideal sinusoidal shape, as well as the structure of the magnon spectrum determined via inelastic 
neutron scattering. In the continuation of this paper we show that the KSEA anisotropy also explains: 
(a) the observed lattice pinning of the spiral propagation vector along the $x=(1,1,0)$ or 
$y=(1,\bar{1},0)$ direction in the absence of an in-plane field, and (b) a remnant of the bisection 
rule observed via neutron diffraction in the presence of an in-plane field \cite{5}.
The KSEA extension of Hamiltonian (\ref{eq:3}) reads
\begin{eqnarray}
\label{eq:14}  
 W  &=& \sum_{mn} \Big[ \, J_{1}\,S^{x}_{m,n}S^{x}_{m+1,n} + J_{2}\,S^{\,y}_{m,n}S^{\,y}_{m+1,n} 
+ \,J_{3}\,S^{\,z}_{m,n}S^{\,z}_{m+1,n} 
\nonumber
\\
& & + \,J_{2}\,S^{\,x}_{m,n}S^{\,x}_{m,n+1} + J_{1}S^{\,y}_{m,n}S^{\,y}_{m,n+1} 
+ \,J_{3}\,S^{\,z}_{m,n}S^{\,z}_{m,n+1} 
\nonumber
\\
& & +\, D_{\perp}\big( \,S^{z}_{m,n}S^{\,x}_{m+1,n} 
- S^{\,x}_{m,n}S^{\,z}_{m+1,n} + S^{\,y}_{m,n}S^{\,z}_{m,n+1} - S^{\,z}_{m,n}S^{\,y}_{m,n+1}\,\big)\, \Big],
\end{eqnarray}
where
\begin{equation}
\label{eq:15}
J_{1} = J - {\Delta}, \hspace{1.2cm} J_{2} = J + {\Delta}, \hspace{1.2cm} J_{3} = J - {\Delta}
\end{equation}
and
\begin{equation}
\label{eq:16}
  {\Delta}\,\, =\,\, \frac{D^{\,2}_{\perp}}{4J}\,\, =\,\, \frac{1}{4}\,{\varepsilon}^{2}J.
\end{equation}
One should keep in mind that Hamiltonian (\ref{eq:14}) is still consistent with the underlying 
space group $P\bar{4}2_{1}m$ but is not the most general Hamiltonian allowed by symmetry \cite{9,10}.
Nevertheless, Eq. (\ref{eq:14}) has been the starting point for a reasonably successful analysis of 
a vast set of experimental data and will be adopted here without further questioning.

Our aim is then to find the classical minima of Hamiltonian (\ref{eq:14}). One may again employ 
the spherical parametrization (\ref{eq:4}) but the resulting coupled equations for ${\Theta}_{m,n}$ and 
${\Phi}_{m,n}$ do not seem to admit an analytical solution (when $\Delta \neq 0$) of the type obtained 
earlier in Eqs. (\ref{eq:5})-(\ref{eq:7}) within the pure DM model ($\Delta=0$). Instead, our strategy 
is to derive a perturbative expansion for small values of $\varepsilon=D_{\perp}/J$ of practical 
interest.

To leading order in $\varepsilon$, the ground state may be obtained by minimizing a suitable continuum 
energy functional \cite{7,9,10}. The resulting spiral generalizes Eq. (\ref{eq:13}) to the extent that 
anharmonic distortions occur in the profile of the angular variable ${\Theta}_{m,n}$ which is no longer 
a linear function 
of ${\varepsilon}(m\cos{\psi} + n\sin{\psi})\approx(x\cos{\psi} + y\sin{\psi})$. 
Instead, the complete solution may be written as
\begin{equation}
\label{eq:17}
  {\Theta}_{m,n} \approx \theta\left(x \, \cos{\psi} + y\, \sin{\psi} \right), 
\hspace{1.0cm} {\Phi}_{m,n} = -\psi ,
\end{equation}
where $\psi$ is an arbitrary constant angle and ${\theta}(u)$ is a function of 
$u = x\cos{\psi} + y\sin{\psi}$ that may be expressed in terms of an elliptic integral. 
This family of solutions is degenerate in the sense that the energy is independent of 
the angle $\psi$. As a consequence, the bisection rule summarized in Fig. 2 remains valid 
in the presence of KSEA anisotropy to leading order in $\varepsilon$.

However, the above degeneracy is again broken beyond the leading approximation. To demonstrate
this fact we have systematically carried out the continuum expansion beyond the leading approximation.
Technical details are too cumbersome to be described here in any detail, but the 
main conclusion may be stated by quoting the final result for the ground-state energy per
unit cell calculated to order ${\varepsilon}^{4}$:
\begin{equation}
\label{eq:18}
  w = {\varepsilon}^{2}s^{2}J\Big\{-\frac{1}{2}{\delta}^{2}
  + \Big[\, C_{0} + C_{1}\cos{(4\psi)}\,\Big]{\varepsilon}^{2} + ...\Big\}.
\end{equation}
When this procedure is applied to the pure DM model ($\Delta=0$) we find that 
${\delta}^{2}=1$, $C_{0}=3/32$, and $C_{1}=1/32$, which agree with the results of Eq. (\ref{eq:12}) 
and thus provide an important check of consistency. In the presence of KSEA anisotropy 
($\Delta={\varepsilon}^{2}J/4$) we find the values 
\begin{eqnarray}
\label{eq:19}
  {\delta}^{2} = 0.53189772, &\hspace{0.8cm} C_{0} = 0.00098724, & \hspace{0.8cm} C_{1} = -0.00474723,
\end{eqnarray}
obtained by numerical calculation of a large number of elliptic-type integrals.
The most important feature of Eq. (\ref{eq:19}) is that the constant $C_{1}$ is now negative 
and thus leads to minimum energy in Eq. (\ref{eq:18}) for $\psi=0$ or $\pi/2$; i.e., for a 
spiral that propagates along the $x$ or $y$ axis, in agreement with experiment.

\begin{figure}
\centerline{\hbox{\psfig{figure=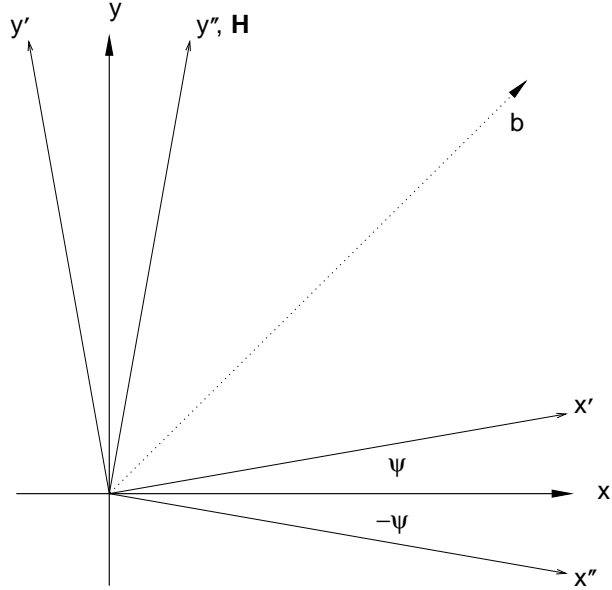,width=8.cm}}}
\vspace*{0.5cm}
\caption{Illustration of the ideal bisection rule. The spiral propagation 
vector points along the $x^{\prime}$ axis, while the spin rotates within 
the $x^{\prime\prime}z$ plane. This ideal rule is experimentally realized 
when a sufficiently strong in-plane magnetic field $\bf{H}$ is applied
along the $y^{\prime\prime}$ axis.}
\end{figure}
The strength of anisotropy appropriate for the description of Ba$_{2}$CuGe$_{2}$O$_{7}$ is estimated
from the value of the spiral pitch measured at zero field \cite{7} to be 
${\varepsilon}=D_{\perp}/J\approx 0.18$. Therefore, anisotropy appears to be sufficiently weak 
for the validity of the leading-order continuum approximation. For instance, the relative correction
induced by the second-order term in Eq. (\ref{eq:18}) is of the order $5\times10^{-4}$. Nevertheless, 
such a tiny correction is important for explaining the correct lattice pinning of the spiral 
propagation vector along the $x$ or $y$ axis, as well as some measurable deviations from the ideal 
bisection rule summarized in Fig. 2.

The bisection rule was initially discovered through neutron diffraction \cite{5} in the presence 
of an in-plane magnetic field
\begin{equation}
\label{eq:20}
 {\bf H} = \left(\cos{\chi}, \sin{\chi}, 0\right) H, 
\end{equation}
which forms an angle $\chi$ with respect to the $x$ axis. The spiral was then observed 
to adjust its propagation vector according to the rule
\begin{equation}
\label{eq:21}
 \chi + \psi = \frac{\pi}{2}, 
\end{equation}
in order to minimize the Zeeman energy induced by the magnetic field. Returning to Fig. 2, 
a spiral that initially propagates along the $x$ axis ($\psi=0$) is reoriented to propagate 
along the $x^{\prime}$ axis ($\psi \neq 0$) so that the normal to the spin plane 
(axis $y^{\prime\prime}$) points along the applied magnetic field.

Actually, the above rule was observed only for sufficiently strong fields $H > 0.5$ T. For 
weak fields, reorientation of the spin structure to conform with the ideal bisection rule is inhibited 
by the small tetragonal anisotropy that is present in the second order term of Eq. (\ref{eq:18}). The 
appropriate modification of Eq. (\ref{eq:21}) is then derived by an argument similar to that given in 
Ref. \cite{5}, now taking into account the theoretical prediction for the tetragonal anisotropy 
contained in Eq. (\ref{eq:18}):
\begin{equation}
\label{eq:22}
H^{2} = \frac{8A\,\sin{(4\psi)}}{\sin{2(\psi+\chi)}}, 
\end{equation}
where
\begin{equation}
\label{eq:23}
 A = \left(\frac{2sJ}{g_{ab}\,{\mu}_{B}}\right)^{2}
\frac{2\,|C_{1}|\,{\varepsilon}^{4}}{\langle \,\sin{^{2}\theta}\,\rangle} 
\end{equation}
is a numerical factor which is here calculated in terms of the constant $C_{1}$ of Eq. (\ref{eq:19}), 
the average ${\langle\sin{^{2}\theta}\,\rangle}=0.56517$ taken over a period of the spiral profile 
of Eq. (\ref{eq:17}), and some basic parameters determined by independent experiments \cite{3,4,5,6,7}; 
namely, the spin value $s=1/2$, the exchange constant $J=0.96$ meV, the DM anisotropy 
$\varepsilon=D_{\perp}/J=0.1774$, the gyromagnetic ratio $g_{ab}=2.044$ for a field applied 
in some direction within the basal plane, and the Bohr magneton ${\mu}_{B}$.

\begin{figure}
\centerline{\hbox{\psfig{figure=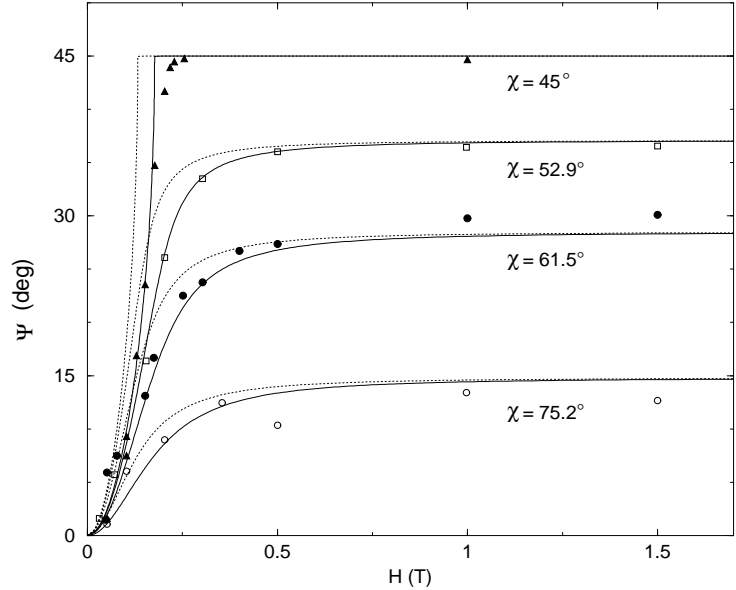,width=10.cm}}}
\vspace*{0.5cm}
\caption{Deviations from the ideal bisection rule predicted by Eq.(\ref{eq:22})
applied for $A=1.1\times10^{-3}$ T$^{2}$ (dotted lines) and 
$A=1.95\times10^{-3}$ T$^{2}$ (solid lines). Note that the ideal bisection
rule ($\chi + \psi = \pi/2$) is indeed realized for a sufficiently strong
in-plane magnetic field $H > 0.5$ T. Experimental data denoted by symbols
were extracted form Fig. 3 of Ref. [5].}
\end{figure}
Putting everything together, we find that $A=1.1\times10^{-3}$ T$^{2}$. The corresponding predictions
of Eq. (\ref{eq:22}) are depicted by dotted lines in Fig. 3 where we display the direction of the 
spiral propagation (angle $\psi$) as a function of applied field ($H$) for various field 
orientations (angle $\chi$). The ideal bisection rule of Eq. (\ref{eq:21}) is clearly reproduced 
for sufficiently strong fields ($H > 0.5$ T) but measurable deviations occur for weak fields. 
Altogether, our theoretical predictions are in reasonable agreement with experiment. A better fit 
is obtained by treating the constant $A$ in Eq. (\ref{eq:22}) as an adjustable phenomenological 
parameter determined by a global fit of the experimental data, as was originally done in Ref. \cite{5} 
to obtain $A=1.95\times10^{-3}$ T$^{2}$. The corresponding predictions of Eq. (\ref{eq:22}) are 
also depicted in Fig. 3 (solid lines) and provide an improved overall fit of the data.

To summarize, the KSEA extension of the pure DM anisotropy is capable of explaining a large 
set of data pertaining to the IC phase transition observed in Ba$_{2}$CuGe$_{2}$O$_{7}$, as well 
as some finer issues such as the lattice pinning of helical magnetic domains. One should 
keep in mind that DM$+$KSEA is not the most general type of anisotropy allowed by 
symmetry \cite{7,9,10}. In this respect, we mention here that a significant departure from KSEA seems
to occur in the layered antiferromagnet K$_{2}$V$_{3}$O$_{8}$. The space group of this compound 
is different ($P4bm$) but leads to a Heisenberg Hamiltonian that is very similar to the 
one encountered in Ba$_{2}$CuGe$_{2}$O$_{7}$. However, an experimental investigation \cite{15} 
of K$_{2}$V$_{3}$O$_{8}$ revealed the occurrence of interesting spin-flop and 
spin-reorientation transitions but provided no evidence for incommensurate magnetism. 
A preliminary theoretical analysis \cite{15,16} concluded that K$_{2}$V$_{3}$O$_{8}$ is characterized 
by an easy-axis anisotropy, which is impossible to occur in the KSEA limit \cite{10}.

\vskip5mm
This paper is dedicated to our friend Professor Victor G. Bar$^{\prime}$yakhtar on the occasion 
of his 75th birthday - helimagnetism has been one of the many subjects of interest during 
his long and fruitful career \cite{17}.





\end{document}